\documentclass[letterpaper,copyedit]{article}   
\usepackage{osajnl2} 
\usepackage{graphicx, amsmath}
\usepackage{cite}
\usepackage{subfigure}
\begin{document}

\title{Preparation of Knill-Lafamme-Milburn states based on superconducting qutrits}


\author{Qi-Gong Liu, Qi-Cheng Wu, and Xin Ji$^{\dag}$}

\address{Department of Physics, College of Science, Yanbian
University, Yanji, Jilin 133002, China}
\address{$^\dag$Corresponding author: jixin@ybu.edu.cn}

\begin{abstract}
We propose two schemes for generating the Knill-Lafamme-Milburn
(KLM) states of two distant polar molecules (PMs) ensembles
respectively in two transmission-line resonators (TLRs) connected
by a superconducting charge qutrit (SCQ), and of two SCQs in a
TLR, respectively. Both of the schemes are robust against photon
decay due to the virtual excitations of the microwave photons of
the TLRs, and the spontaneous emission can be suppressed owing to
the virtual transitions of the SCQs in the second scheme. In
addition, the schemes have high controllability and feasibility
under the current available techniques.
\end{abstract}

\maketitle 

\noindent {\bf 1. INTRODUCTION}

Quantum entanglement, an interesting and attractive phenomenon in
quantum mechanics, plays a significant role not only in testing
quantum nonlocality, but also in processing a variety of quantum
information tasks \cite{AKE,CHB,CHBS,CHBG,KMH,MHVB,SBZGCG,GV}.
Therefore, preparation of various quantum entangled states has
been being an important subject in quantum information science
since a few decades ago
\cite{EHX,SB,GCGY,PBSO,DLEK,JCH,XBZ,CYL,DBH,XQS,HYYL,XYCPY}. In
2001, Knill, Lafamme and Milburn proposed a specific class of
partially entangled states \cite{EK}, KLM states, and they derived
that employing the KLM states as ancillary resources can improve
the success probability of teleportation gradually upto unity with
the increase of the particle number in these ancillary states, and
thereafter, Mod\l awska and Grudka showed that multiple adaptive
teleportation in the KLM scheme can also increase the probability
of faithful teleportation \cite{JMAG}, which can elevate the
efficiency of quantum computing significantly. In the past dozen
years, the investigation of the KLM states preparation has
attracted a great deal of attention\cite{JD,KLJF,KLAC,KL,LYC}. The
first explicit scheme to prepare the KLM states was proposed by
Franson \emph{et al.} in 2004 by using elementary linear optics
gates and solid-state approach, respectively \cite{JD}. In 2008,
Lemr \emph{et al.} proposed a scheme to prepare the two-photon
four-mode KLM states using linear optical elements \cite{KLJF}.
Soon afterwards, preparing high fidelity two-photon KLM states
experimentally was implemented using spontaneous parametric
down-conversion photon source and linear optical components in
2010 \cite{KLAC}. In 2011, Lemr proposed a scheme to prepare KLM
states by using a tunable controlled phase gate and optimized the
scheme for the framework of liner optics \cite{KL}. In 2012, Cheng
\emph{et al}. proposed two schemes to prepare the two-atom KLM
states with a strong coupling cavity-fiber system and the
cavity-assisted single-photon input-output process, respectively
\cite{LYC}.

Although it has been verified that photonic qubits and atomic
qubits can be used to realize the KLM-type quantum computation
\cite{SP}, single-photon detectors needed for photon KLM
computation are inefficient due to photon losses or dark counts,
while the manipulation of the atom is still a severe challenge in
the state of the art though great progress has been made in recent
decades. So it is hoped to find an alternative candidate of photon
and atom, which not only has high usability and feasibility, but
also has better controllability and enormous superiority.
Fortunately, the SCQ concerning the interaction with the TLR in
microwave cavities provides a promising candidate of the physical
system for quantum information processing. In this paper, we put
forward two schemes to generate the KLM states of two distant PMs
ensembles via a hybrid device and of two SCQs in a TLR,
respectively. Our schemes have following advantages: 1) With the
help of the SCQ and TLR, our schemes have longer coherence time
and storage time than that in Refs.~\cite{JD,KLJF,KLAC,KL,LYC}. 2)
Compared with the Ref. \cite{LYC}, it is needless to take the
coding qubits out of the TLR in our schemes, which avoids the
decoherence induced by the environment. 3) The controllability and
feasibility of our schemes are higher than that in
Refs.~\cite{JD,KLJF,KLAC,KL,LYC} in the current techniques,
because the strong coupling in our schemes can be obtained by
increasing the number of PMs in each ensemble \cite{QC,PR} in the
first scheme or locating the SCQs at the antinodes of the voltage
\cite{AB} in the second scheme. 4) Both of the schemes are robust
against photon decay due to the virtual excitations of the
microwave photons of the TLRs, and the spontaneous emission can be
suppressed owing to the virtual transitions of the SCQs in the
second scheme. 5) The second scheme can be generalized to prepare
$N$ qubits KLM states straightforward.

The rest of the present paper is organized as follows. In Section
2, we give a scheme to prepare the two qubits KLM states of two
PMs, then in Section 3 we give another scheme to prepare the two
qubits KLM states of two SCQs and generalize the scheme to $N$
qubits. The feasibility analysis and conclusion are given in
Section 4.

\noindent {\bf 2. PREPARATION OF THE KLM STATES OF DISTANT POLAR
MOLECULES ENSEMBLES VIA A HYBRID DEVICE}

We consider a hybrid device with two cold PMs ensembles, two TLRs
and a SCQ, as shown in Fig.~1. Here, the SCQ capacitively coupling
to the two TLRs can be viewed as an artificial three-level atom
with two stable ground states $|i\rangle_{s},~|g\rangle_{s}$ and
an excited state $|e\rangle_{s}$ \cite{CPY,CPY2}. The transition
between $|g\rangle_{s}$ and $|e\rangle_{s}$ with frequency
$\omega_{s}$ is dispersively coupled to the TLR mode. The cold PMs
ensembles are placed respectively at the antinodes of the two TLRs
and possess stable rotational states and can be well controlled by
microwave fields \cite{LDC,SDH,BZ}. When the PMs are cooled to the
ground states and vibrational states of electrons, the PMs can be
viewed as two-level system with excited state $|e\rangle_{m}$ and
ground state $|g\rangle_{m}$ under external weak electric fields.
Assume that there are $N$ identical PMs in each PMs ensemble and
they have no interaction each other. Under driven by a classical
microwave field with frequency $\omega_{d}$ and Rabi frequency
$\Omega$, the whole dynamics of the combined system involving a
cold PMs ensemble, a TLR and a SCQ are governed by the Hamiltonian
$(~\hbar=1~)$ \cite{QC,MFC}
\begin{equation}\label{01}
H_{s}=\frac{\omega_{s}}{2}\sigma^{z}+\Omega(\sigma^{-}e^{i\omega_{d}t}+\sigma^{+}e^{-i\omega_{d}t})+\frac{\omega_{m}}{2}S^{z}+\omega_{c}a^{\dagger}a+g_{s}(\sigma^{+}a+\sigma^{-}a^{\dagger})+g_{m}(S^{+}a+S^{-}a^{\dagger}),
\end{equation}
where
$\sigma^{z}=|e\rangle_{s}\langle{e}|-|g\rangle_{s}\langle{g}|$,
$\sigma^{+}=|e\rangle_{s}\langle{g}|,~\sigma^{-}=|g\rangle_{s}\langle{e}|
$.
$S^{z}=\sum^{N}_{j=1}s^{z}_{j}(s^{z}_{j}=|e\rangle_{m,j}\langle{e}|-|g\rangle_{m,j}\langle{g}|
)$,
$S^{+}=\sum^{N}_{j=1}s^{+}_{j}(s^{+}_{j}=|e\rangle_{m,j}\langle
g|)$ and
$S^{-}=\sum^{N}_{j=1}s^{-}_{j}(s^{-}_{j}=|g\rangle_{m,j}\langle{e}|)$
are the collective spin operators for the PMs. $a^{\dagger}$ and
$a$ are the creation and annihilation operators of the microwave
photon with frequency $\omega_{c}$. $g_{s}(g_{m})$ is the coupling
strength between the SCQ (PMs) and the TLR. Under the conditions
$|\Delta_{s}|=|\omega_{s}-\omega_{c}| \gg g_{s} $, $ |\Delta_{
m}|= |\omega_{m}-\omega_{c}| \gg g_{m} $ and $|\Delta_{d}|=
|\omega_{s}-\omega_{d}| \gg \Omega$, the dispersive interaction
and classical field induce the Stark shifts, respectively. Letting
$b=(1/\sqrt{N})S^{-},~b^{\dagger}=(1/\sqrt{N})S^{+}$ and
$n_{b}=\sum^{N}_{j=1}|e\rangle_{m,j}\langle e|$, then
$[b,b^{\dagger}]=1-(2/N)n_{b},~[n_{b},b^{\dagger}]=b^{\dagger}$
and $[n_{b},b]=-b $. In the low-excitation case, $b^{\dagger}$ and
$b$ can be regarded as the bosonic operators and the PMs ensemble
can be regarded as a bosonic system.  If the TLR is initially in
the vacuum state, the Hamiltonian $H_{s}$ can be reduced to
\cite{MFC}
\begin{equation}\label{02}
H_{eff}=\frac{1}{2}(2\lambda_{sd}+\lambda_{sc})\sigma^{z}+g(\sigma^{+}b+\sigma^{-}b^{\dagger})+N\lambda_{mc}b^{\dagger}b,
\end{equation}
where $\lambda_{sd}=\Omega^{2}/\Delta_{d},~
\lambda_{sc}=g_{s}^{2}/\Delta_{s},~
\lambda_{mc}=g_{m}^{2}/\Delta_{m}$, $g=\sqrt{N}\lambda_{sm}~(
\lambda_{sm}=g_{m}g_{s}/2(1/\Delta_{m}+1/\Delta_{s}))$ and the
term of the constant energy has been discarded. Under the resonant
condition $2\lambda_{sd}+\lambda_{sc}=N\lambda_{mc}$, the
effective Hamiltonian $H_{eff}$ leads to the resonant interaction
between the SCQ and the PMs ensemble. We can have the following
transitions:
\begin{equation}\label{03}
|e\rangle_{s}|n\rangle\rightarrow\cos(g\sqrt{n+1}t)|e\rangle_{s}|n\rangle-i\sin(g\sqrt{n+1}t)|g\rangle_{s}|n+1\rangle,
\end{equation}
\begin{equation}\label{04}
~~~~|g\rangle_{s}|n+1\rangle\rightarrow\cos(g\sqrt{n+1}t)|g\rangle_{s}|n+1\rangle-i\sin(g\sqrt{n+1}t)|e\rangle_{s}|n\rangle,
\end{equation}
where $|n\rangle$ represents the $n$-excitation Fock state of the
PMs mode. Here the common phase factor is discarded.

Now the method for preparing the KLM states of the collective
modes of two distant PMs ensembles is given based on the resonant
interaction discussed above. Assume that the SCQ is initially in
the state $|i\rangle_{s}$ and the two PMs ensembles are initially
in the vacuum state $|0\rangle_{1}|0\rangle_{2}$, where the
subscript 1 and 2 indicate different PMs ensembles. The operations
for realizing the KLM states are described as below:

$Step$ 1: Adjust the energy level gap of the SCQ \cite{JC,MN,JQY}
so that the SCQ does not interact with two PMs ensembles and apply
a classical microwave pulse with the Rabi frequency $ \Omega^{'} $
to the SCQ, then the transition
$|i\rangle_{s}\rightarrow\cos{({\Omega^{'}}t_{0})}|i\rangle_{s}-ie^{-i\phi}\sin{({\Omega^{'}}t_{0})}|e\rangle_{s}$
will be performed. Hence the state of the combined system is
\begin{equation}\label{05}
|\Psi\rangle_{1}=[\cos{({\Omega^{'}}t_{0})}|i\rangle_{s}-ie^{-i\phi}\sin{({\Omega^{'}}t_{0})}|e\rangle_{s}]|0\rangle_{1}|0\rangle_{2}.
\end{equation}
~~~~$Step$ 2: Let the SCQ interact with ensemble 1 and set the
interaction time $t_{1}=\pi/(2g)$. The state of the combined
system evolves to
\begin{equation}\label{06}
|\Psi\rangle_{2}=[\cos{({\Omega^{'}}t_{0})}|i\rangle_{s}|0\rangle_{1}-e^{-i\phi}\sin{({\Omega^{'}}t_{0})}|g\rangle_{s}|1\rangle_{1}]|0\rangle_{2}.
\end{equation}
~~~~$Step$ 3: Adjust the energy level gap of the SCQ again to
decouple the SCQ from the TLRs and apply another classical
microwave pulse with the Rabi frequency $\Omega^{''}$ to the SCQ
for implementing the transition
$|g\rangle_{s}\rightarrow\cos{(\Omega^{''}t_{2})}|g\rangle_{s}-ie^{-i\phi^{'}}\sin{(\Omega^{''}t_{2})}|e\rangle_{s}$.
Thus the state of the combined system evolves into
\begin{eqnarray}\label{07}
|\Psi\rangle_{3}&=&\cos{({\Omega^{'}}t_{0})}|i\rangle_{s}|0\rangle_{1}|0\rangle_{2}-e^{-i\phi}\sin{({\Omega^{'}}t_{0})}\cos{(\Omega^{''}t_{2})}|g\rangle_{s}|1\rangle_{1}|0\rangle_{2}\cr
&&+ie^{-i(\phi+\phi^{'})}\sin{({\Omega^{'}}t_{0})}\sin{(\Omega^{''}t_{2})}|e\rangle_{s}|1\rangle_{1}|0\rangle_{2}.
\end{eqnarray}
~~~~$Step$ 4: Let the SCQ interact with ensemble 2 and choose the
interaction time $t_{3}=\pi/(2g)$, after that the two collective
modes of two PMs ensembles evolve to
\begin{eqnarray}\label{08}
|\Psi\rangle_{4}&=&\cos{({\Omega^{'}}t_{0})}|i\rangle_{s}|0\rangle_{1}|0\rangle_{2}-e^{-i\phi}\sin{({\Omega^{'}}t_{0})}\cos{(\Omega^{''}t_{2})}|g\rangle_{s}|1\rangle_{1}|0\rangle_{2}\cr
&&+e^{-i(\phi+\phi^{'})}\sin{({\Omega^{'}}t_{0})}\sin{(\Omega^{''}t_{2})}|g\rangle_{s}|1\rangle_{1}|1\rangle_{2}.
\end{eqnarray}
~~~~$Step$ 5: Apply a classical microwave pulse to the SCQ for the
transition $|i\rangle_{s}\rightarrow|e\rangle_{s}$, so the state
becomes
\begin{eqnarray}\label{9}
|\Psi\rangle_{5}&=&\cos{({\Omega^{'}}t_{0})}|e\rangle_{s}|0\rangle_{1}|0\rangle_{2}-e^{-i\phi}\sin{({\Omega^{'}}t_{0})}\cos{(\Omega^{''}t_{2})}|g\rangle_{s}|1\rangle_{1}|0\rangle_{2}\cr
&&+e^{-i(\phi+\phi^{'})}\sin{({\Omega^{'}}t_{0})}\sin{(\Omega^{''}t_{2})}|g\rangle_{s}|1\rangle_{1}|1\rangle_{2}.
\end{eqnarray}
~~~~$Step$ 6: Apply a classical microwave pulse to the SCQ again
for the transitions
$|g\rangle_{s}\rightarrow1/\sqrt{2}(|g\rangle_{s}-|e\rangle_{s})$,
$|e\rangle_{s}\rightarrow1/\sqrt{2}(|e\rangle_{s}+|g\rangle_{s})$.
So the final state of the combined system turns into
\begin{eqnarray}\label{10}
|\Psi\rangle_{6}&=&\frac{1}{\sqrt{2}}|g\rangle_{s}[\alpha|0\rangle_{1}|0\rangle_{2}-\beta|1\rangle_{1}|0\rangle_{2}+\gamma|1\rangle_{1}|1\rangle_{2}]\cr
&&+\frac{1}{\sqrt{2}}|e\rangle_{s}[\alpha|0\rangle_{1}|0\rangle_{2}+\beta|1\rangle_{1}|0\rangle_{2}-\gamma|1\rangle_{1}|1\rangle_{2}],
\end{eqnarray}
where
$\alpha=\cos{({\Omega^{'}}t_{0})},~\beta=e^{-i\phi}\sin{(\Omega^{'}{t_{0}})}\cos{(\Omega^{''}t_{2})},~\gamma=e^{-i(\phi+\phi^{'})}\sin{({\Omega^{'}}t_{0})}\sin{(\Omega^{''}t_{2})}$.
If $\alpha=\beta=\gamma=1/\sqrt{3}$ are set, i.e.,
${\Omega^{'}}t_{0}=\arccos(1/\sqrt{3}),~\Omega^{''}t_{2}= \pi/4$
and $\phi=\phi^{'}=0$, the state becomes
\begin{eqnarray}\label{11}
|\Psi\rangle_{7}&=&\frac{1}{\sqrt{6}}|g\rangle_{s}[|0\rangle_{1}|0\rangle_{2}-|1\rangle_{1}|0\rangle_{2}+|1\rangle_{1}|1\rangle_{2}]\cr
&&+\frac{1}{\sqrt{6}}|e\rangle_{s}[|0\rangle_{1}|0\rangle_{2}+|1\rangle_{1}|0\rangle_{2}-|1\rangle_{1}|1\rangle_{2}].
\end{eqnarray}
Then, a detection on the SCQ should be performed. It is worth
noting that no matter what the measurement result is, the KLM
states can be always achieved. That's to say, the successful
probability of our protocol is unity in the ideal case. In
addition, according to the measurement result, a simple local
operation $\sigma^{z}$ can be performed on the first PMs ensemble
via a feedback process to achieve the same KLM states.

Up to now, all the above results are based on the ideal case. In
fact, in our scheme, a series of interacting times should be
controlled accurately to prepare the KLM states with a high
fidelity. So the effect of the time errors on the fidelity should
be discussed. Here, the fidelity is defined as
$F=|_{6}\langle{\Psi}|{\varphi}\rangle_{6}|^{2}$, where
$|{\varphi}\rangle_{6}$ is the final state of system when the time
errors $\Delta t_{j}~ (j=0,1,2,3)$ are introduced (It should be
noted that we have ignored the the time errors in \emph{Steps} 5
and 6, because the precise control of microwave pulse has already
been shown by several groups~\cite{GZS,UK}). After calculation we
find $F$ can be evaluated as
\begin{eqnarray}\label{12}
F=|\alpha^{\ast}\alpha^{'}+\beta^{\ast}\beta^{'}+\gamma^{\ast}\gamma^{'}|^2,
\end{eqnarray}
where~$\alpha^{'}=\cos[{{\Omega^{'}}(t_{0}+\Delta
t_{0})}],~\beta^{'}=e^{-i\phi}\sin{[{\Omega^{'}}(t_{0}+\Delta
t_{0})]}\sin{[g(t_{1}+\Delta
t_{1})]}\cos{[{\Omega^{''}}(t_{2}+\Delta
t_{2})]},~\gamma^{'}=e^{-i(\phi+\phi^{'})}\sin[{{\Omega^{'}}(t_{0}+\Delta
t_{0})}]\sin{[g(t_{1}+\Delta
t_{1})]}\sin{[{\Omega^{''}}(t_{2}+\Delta
t_{2})]}\sin{[g(t_{3}+\Delta t_{3})]}$. In order to discuss the
effect of time errors on the fidelity, we still let
${\Omega^{'}}t_{0}=\arccos(1/\sqrt{3})$, $\Omega^{''}t_{2}=
\pi/4$, $gt_{1}=gt_{3}=\pi/2$ as above, and define the time error
rate as $\eta_{j}=\Delta t_{j}/t_{j}$ ($j=0,1,2,3$) (Here,
$\eta_{j} (j=0,1,2,3)$ is the relative error of time, $t_{1}$
($t_{3}$) the interaction time between SCQ and PMs 1 (PMs 2),
$t_{0}$ ($t_{2}$) the operation time of classical pulse with Rabi
frequency $\Omega^{'}$ ($\Omega^{''}$)). Density plots of the
fidelity as a function of $\eta_{1}$ and $\eta_{3}$ are shown in
Fig.~2. It can be seen from Fig.~2(a) the fidelity decreases
slightly with the increase of $\eta_{1}$ and $\eta_{3}$ when
$\eta_{0}=\eta_{2}=0$, and the fidelity can still reach $0.975$
when $\eta_{1}=\eta_{3}=0.1$. Fig.~2(b) shows that the fidelity is
still as high as 0.96 when $\eta_{1}=\eta_{3}=0.1$ even with
$\eta_{0}=\eta_{2}=0.1$.

\noindent {\bf 3. PREPARATION OF THE KLM STATES USING A TUNABLE
CONTROLLED PHASE GATE OF SUPERCONDUCTING QUTRITS}

In this section, we propose another scheme to prepare the KLM
states by using a tunable controlled phase gate of superconducting
circuits. The experimental device involving two SCQs capacitively
coupling to a TLR for implementing the controlled phase gate and
the level configuration of the SCQs are shown in Fig.~3. The SCQ
has three levels~$|i\rangle$, $|g\rangle$ and $|e\rangle$, among
which the two stable ground states $|i\rangle$ and $|g\rangle$ are
used to encode qubit information, and the excited state
$|e\rangle$ is used for virtual transitions. The TLR with
eigenfrequency $\omega_{c}$ is strongly detuned from SCQ resonance
by $\Delta=\omega_{c}-(\omega_{e}-\omega_{g})$ and differentially
detuned from the classical driving field by
$\delta=\omega_{d}-\omega_{c}$, as illustrated in Fig.~3(b). The
SCQs are far apart from each other so that the interaction between
SCQs can be ignored \cite{OG}. Then the Hamiltonian of the system
is~\cite{XXY}
\begin{equation}\label{13}
H=\hbar\sum_{{j=1,2}}\left[\left(\omega_{e}-i\frac{\Gamma_{j}}{2}\right)|e\rangle_{j}\langle{e}|+\omega_{g}|g\rangle_{j}\langle{g}|\right]+\frac{1}{2}\sum_{{j=1,2}}(\Omega_{j}\sigma_{j}^{+}e^{-i\omega_{d}t}+g_{j}\sigma_{j}^{+}a+{\rm
{h.c}})+\hbar(\omega_{c}-i\kappa)a^{\dagger}a,
\end{equation}
where
$\sigma^{+}_{j=1,2}=|e\rangle_{j}\langle{g}|,~\sigma^{-}_{j=1,2}=|g\rangle_{j}\langle{e}|$
are the $j$-th spin operators. $a^{\dagger}$ and $a$ are the
creation and annihilation operators of the microwave
photon.~$\Omega_{j}$ is the Rabi frequency of the external
microwave driving field and $g_{j}$ is the coupling strength
between the TLR and SCQs. $\Gamma_{j}$ and $\kappa$ are the decay
rates of SCQs and TLR respectively. By combining the coupling of
the TLR-SCQ and the effect of a classic microwave driving field,
the system will create a dynamical Stark effect for the state
$|g,g\rangle_{12}$, meanwhile, keep the states $|i,i\rangle_{12}$,
$|i,g\rangle_{12}$ and $|g,i\rangle_{12}$ unchanged. That is to
say, the conditional phase $(\varphi)$ gate between SCQ1 and SCQ2
can be implemented under the action of Hamiltonian Eq. (\ref{13})
as follows \cite{XXY}:
\begin{eqnarray}\label{14}
&&|i\rangle_{1}|i\rangle_{2}\rightarrow|i\rangle_{1}|i\rangle_{2},~~~~~|i\rangle_{1}|g\rangle_{2}\rightarrow|i\rangle_{1}|g\rangle_{2},\cr
&&|g\rangle_{1}|i\rangle_{2}\rightarrow|g\rangle_{1}|i\rangle_{2},~~~~|g\rangle_{1}|g\rangle_{2}\rightarrow{e}^{i\varphi}|g\rangle_{1}|g\rangle_{2},
\end{eqnarray}
where the phase $\varphi=-t|\Omega|^2/2(\Delta+\delta)$ is
tunable. The successful realization of the conditional phase gate
is based on the following limits: 1)~$|\Delta|\gg\Gamma_{j},
\kappa, |\Omega_{j}|, |g_{j}|$, and $|\delta|\sim0;~2)~
|g_{j}|>|\Omega_{j}|$ and $|g_{j}|^{2}\gg\Gamma_{j}\kappa$. In the
following calculation, we set the phase $\varphi=-3\pi/2$ which
can be implemented by selecting the appropriate parameters.

Now the KLM states can be prepared using the controlled phase gate
based on the above device. It is assumed that the two SCQs are in
the steady state $|i\rangle_{1}|i\rangle_{2}$. We apply a
classical microwave pulse to SCQ1, SCQ2 respectively to let the
two SCQs in the state
$1/2(|i\rangle_{1}+|g\rangle_{1})(|i\rangle_{2}+|g\rangle_{2})$.
Subsequently, the interactions between the SCQs and TLR are turned
on. Undergoing an appropriate dynamic evolution time, only the
state $|g\rangle_{1}|g\rangle_{2}$ will produce a phase
$\varphi=-3\pi/2$ and the rest states remain unchanged. The state
evolution process is
\begin{equation}\label{15}
\frac{1}{2}(|i\rangle_{1}+|g\rangle_{1})(|i\rangle_{2}+|g\rangle_{2})\rightarrow\frac{1}{2}(|i,i\rangle_{12}+|i,g\rangle_{12}+|g,i\rangle_{12}+i|g,g\rangle_{12}).
\end{equation}
~~~~~Then apply a classical microwave pulse to SCQ2 for the
transformations
$|i\rangle_{2}\rightarrow1/\sqrt{2}(|i\rangle_{2}-|g\rangle_{2}),~|g\rangle_{2}\rightarrow1/\sqrt{2}(|i\rangle_{2}+|g\rangle_{2})$.
The process is
\begin{equation}\label{16}
\frac{1}{2}(|i,i\rangle_{12}+|i,g\rangle_{12}+|g,i\rangle_{12}+i|g,g\rangle)_{12}\rightarrow\frac{1}{2\sqrt{2}}[~2~|i,i\rangle_{12}+(1+i)|g,i\rangle_{12}+(i-1)|g,g\rangle_{12}~].
\end{equation}

Up to now, we have implemented the preparation of the two-qubit
KLM states. Moreover, the scheme can be generalized to generate
$N$-qubit KLM states straightforward. Assuming that the $N-1$
qubits have been in the KLM states and the $N$-th SCQ is in the
state $|i\rangle_{N}$ initially, then the procedure for generating
the $N$-qubit KLM states is as follows:

$Step$ 1: Adjust the previous ($N$-2) SCQs to decouple from the
TLR \cite{JC,MN,JQY} so that the interaction occurs only between
($N$-1)-th and $N$-th SCQs.

$Step$ 2: Apply a classical microwave pulse to $N$-th SCQ and
implement the transform
$|i\rangle_{N}\rightarrow1/\sqrt{2}(|i\rangle_{N}+|g\rangle_{N})$,
as above.

$Step$ 3: Undergoing an appropriate dynamic evolution time, the
state $|g\rangle_{N-1}|g\rangle_{N}$ will produce a phase
$\varphi=-3\pi/2$ and the rest of the states will remain
unchanged.

$Step$ 4: Apply a classical microwave pulse to the $N$-th SCQ once
again for implementing the transforms
$|i\rangle_{N}\rightarrow1/\sqrt{2}(|i\rangle_{N}-|g\rangle_{N}),~~|g\rangle_{N}\rightarrow1/\sqrt{2}(|i\rangle_{N}+|g\rangle_{N})$.
Then, the KLM states of $N$ qubits can be prepared successfully.
Here, we use the notation $|i\rangle^{N}(|g\rangle^{N})$ to denote
the number of the SCQs in the state $|i\rangle(|g\rangle)$.

If $N$ is an even number, the KLM states we prepare is
\begin{equation}\label{17}
|\Psi\rangle_{KLM}=\frac{1}{\sqrt{2}^{N+1}}[\alpha_{0}|i\rangle^{N}+\sum_{{j=1}}^{{N-1}}\alpha_{j}|g\rangle^{{j}}|i\rangle^{{N-j}}+\alpha_{N}|g\rangle^{N}],
\end{equation}
where~$\alpha_{0}=2^{N/2},~\alpha_{j}=-2^{N/2-j+1}(i-1)^{j-2},~\alpha_{N}=-2^{2-N/2}i(i-1)^{N-3}$.

If $N$ is an odd number, the KLM states we prepare is
\begin{equation}\label{18}
|\Psi\rangle_{KLM}=\frac{1}{\sqrt{2}^{N}}[\alpha_{0}|i\rangle^{N}+\sum_{{j=1}}^{{N-1}}\alpha_{j}|g\rangle^{{j}}|i\rangle^{{N-j}}+\alpha_{N}|g\rangle^{N}],
\end{equation}
where~$\alpha_{0}=2^{(N-1)/2},~\alpha_{j}=-2^{(N+1)/2-j}(i-1)^{j-2},~\alpha_{N}=-2^{(3-N)/2}i(i-1)^{N-3}$.

\noindent {\bf 4. THE FEASIBILITY ANALYSIS AND CONCLUSION}

Now we discuss briefly the feasibility of the present schemes
based on the current available parameters. For the first scheme,
we choose the coupling strength between the SCQ and the TLR as $
g_{s}=2\pi\times75$ MHz \cite{LD}, and the coupling strength of
molecule-cavity as $g_{m}=2\pi\times20$ MHz \cite{AA}. If we set
$\Delta_{s}=10g_{s}=2\pi\times750$ MHz,
$\Delta_{m}=25g_{m}=2\pi\times0.5$ GHz and $N=10^{4}$, then the
effective coupling strength between the PMs and the SCQ can reach
as $g=2\pi\times250$ MHz. Furthermore, the effective coupling $g$
can be enhanced greatly with the number $N$ of the PMs of each
ensemble increasing. For the choice of the energy relaxation and
dephasing times of the SCQ as $\tau_{e}=25 $ $\mu$s and $\tau=5 $
$\mu$s \cite{CPY2}, the dephasing rate
$\gamma=1/\tau=2\pi\times0.032$ MHz, and the single-molecule
collision rate is demonstrated to be $\gamma_{\rm
m}\leq2\pi\times700$ Hz \cite{PR}, so the decoherence rates
induced by the SCQ and the PMs are much smaller than the effective
coupling strength. Additionally, the decoherence caused by the
decay of the TLRs could be suppressed effectively due to the
virtual excitations of microwave photons. When we set
$\Omega^{'}=\Omega^{''}=0.1g$, the total time used to generate the
two-qubit KLM states is less than $29$ ns which is far less than
the coherence time of the SCQ.

For the second scheme, the strong coupling between the SCQ and the
TLR can be yielded by fabricating two SCQs at the maximum of the
voltage standing wave~\cite{AB}. Based on currently available
technology, the parameters
$\Delta,~g_{j},~\Omega_{j},~\kappa,~\Gamma_{j}$
$\sim{2}\pi\times(400,~75,~30,~0.008,~0.0064)$ MHz can be
achieved~\cite{CPY,CPY2}. Obviously, the ratio of $\kappa$
($\Gamma_{j}$) to $g_{j}$ is so small that we can safely ignore
the effect of the decay rates of TLR (SCQ) (For simplification of
analysis, we have assumed $\Gamma_{j}=\Gamma$ and $g_{j}=g$).
Furthermore, in our scheme, because the decay rates of the TLR and
the SCQs are smaller than that of the system in Ref.~\cite{XXY},
the conditions for implementing the controlled phase gate are
satisfied more easily, so the success probability of the
controlled phase gate on SCQs is near determinacy. In addition, we
can calculate the gate operation time is about 0.666 $\mu s$ and
the total times used to generate two-, $N$-qubit KLM states are
about 0.674 $\mu s$ and 0.674($N-1$) $\mu s$ respectively.

In conclusion, we have proposed two different methods to prepare
two- and $N(N\geq2)$-qubit KLM states. We have discussed the
feasibilities of the two schemes in detail. By using SCQ, TLR and
encoding information on the two lower levels of the SCQ, our
schemes have longer coherence time and storage time. The
controllability and feasibility of our schemes are high in the
current techniques. Both of the schemes are robust against photon
decay due to the virtual excitations of the microwave photons of
the TLRs, and the spontaneous emission can be suppressed owing to
the virtual transitions of the SCQs' internal states in the second
scheme. Furthermore, we have generalized the second scheme to
prepare $N$-qubit KLM states. We hope that our work may be useful
for the quantum information in the near future.

\begin{center}
{\bf{ACKNOWLEDGMENT}}
\end{center}

This work was supported by the National Natural Science Foundation
of China under Grant Nos. 11064016 and 61068001.

\newpage

\clearpage Fig.~1. Schematic setup of two PMs ensembles placed in
the two separate TLRs coupled by a SCQ and the level diagram for
the PM and the SCQ, where $g_{s}$ is the coupling strength between
TLR and SCQ. $\Omega$ is the rabi frequencie of microwave pulse.

Fig.~2. Density plots of the fidelity as a function of $\eta_{1}$
and $\eta_{3}$ with (a) $\eta_{0}=\eta_{2}=0$. (b)
$\eta_{0}=\eta_{2}=0.1$.

Fig.~3. (a) The schematic circuit of the tunable controlled phase
gate. The setup involves two SCQs and a TLR. (b) The level
configuration of the SCQ.

\clearpage
\begin{figure}
\scalebox{0.9}{\includegraphics{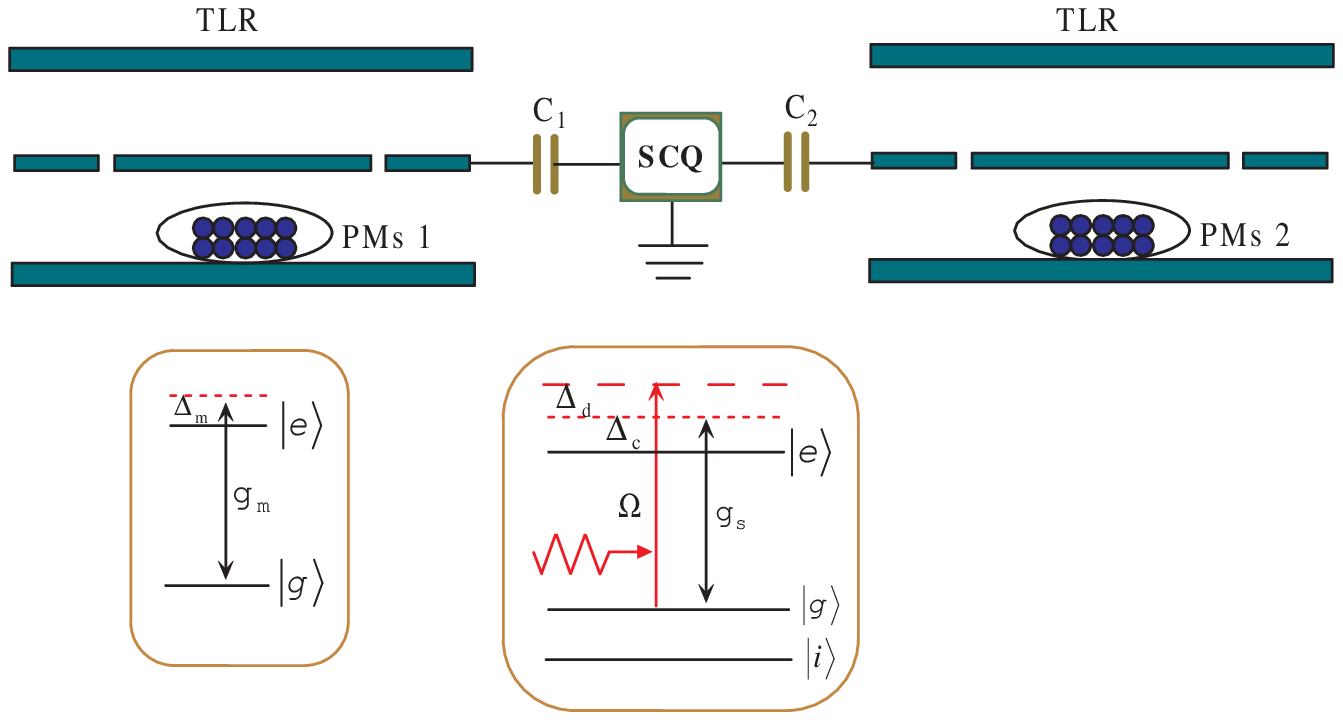}}\caption{\label{f1}}
\end{figure}

\clearpage
\begin{figure}
  \subfigure[]{
    \label{fig2a}
    \includegraphics[width=3.2in,height=2.8in]{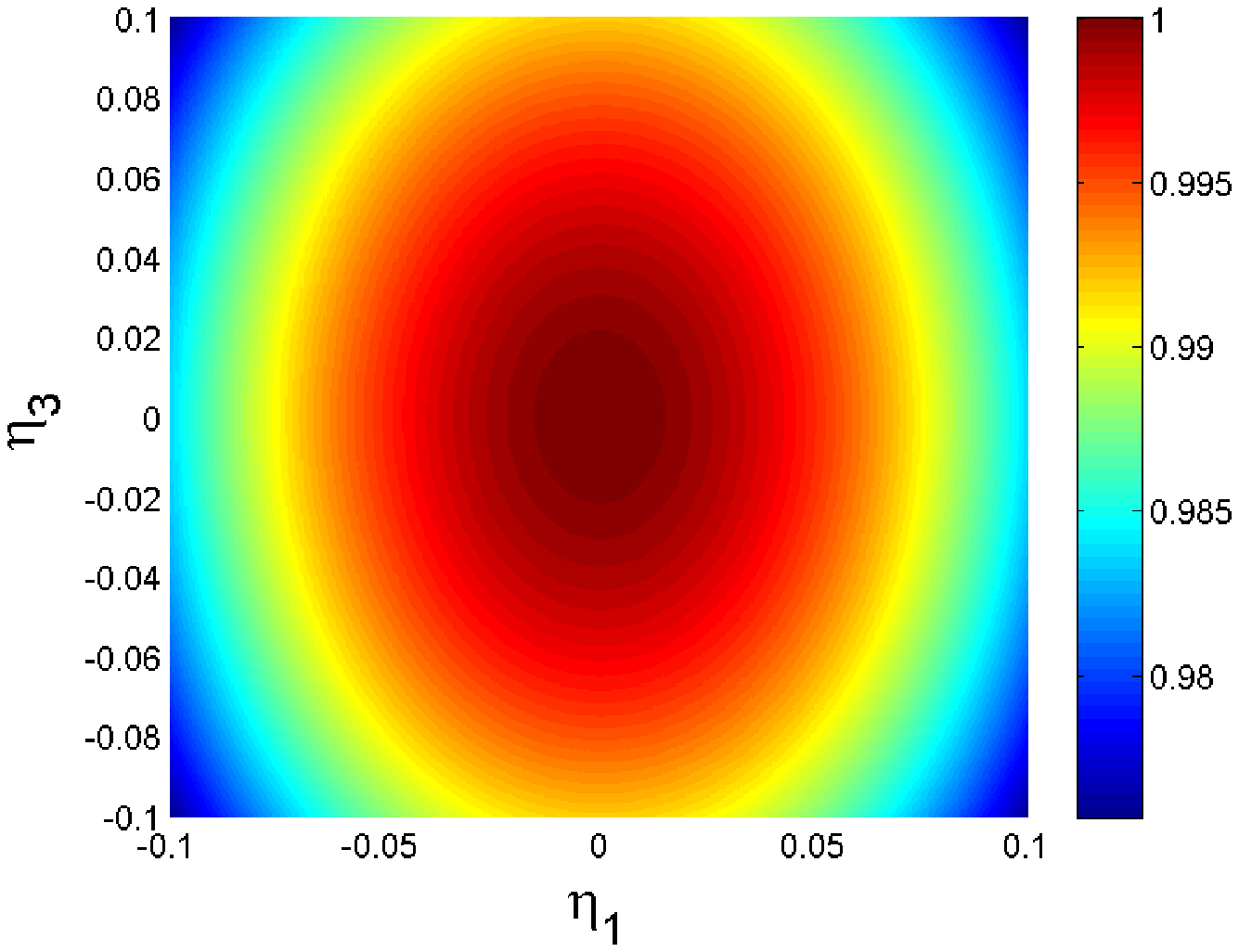}}
  \subfigure[]{
    \label{fig2b}
    \includegraphics[width=3.2in,height=2.8in]{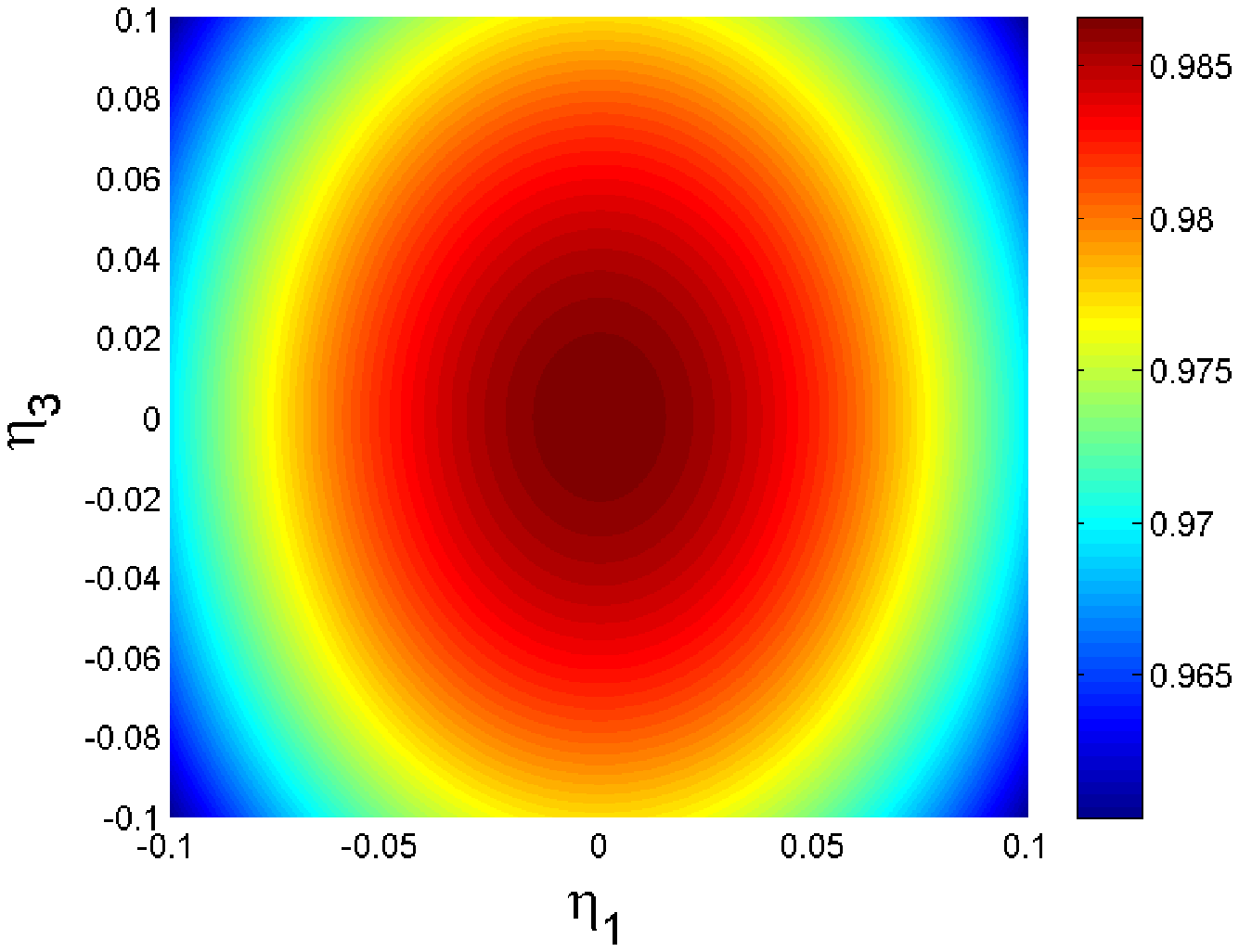}}
 \label{fig.2}
\caption{\label{f2}}
\end{figure}

\clearpage
\begin{figure}
\scalebox{0.9}{\includegraphics{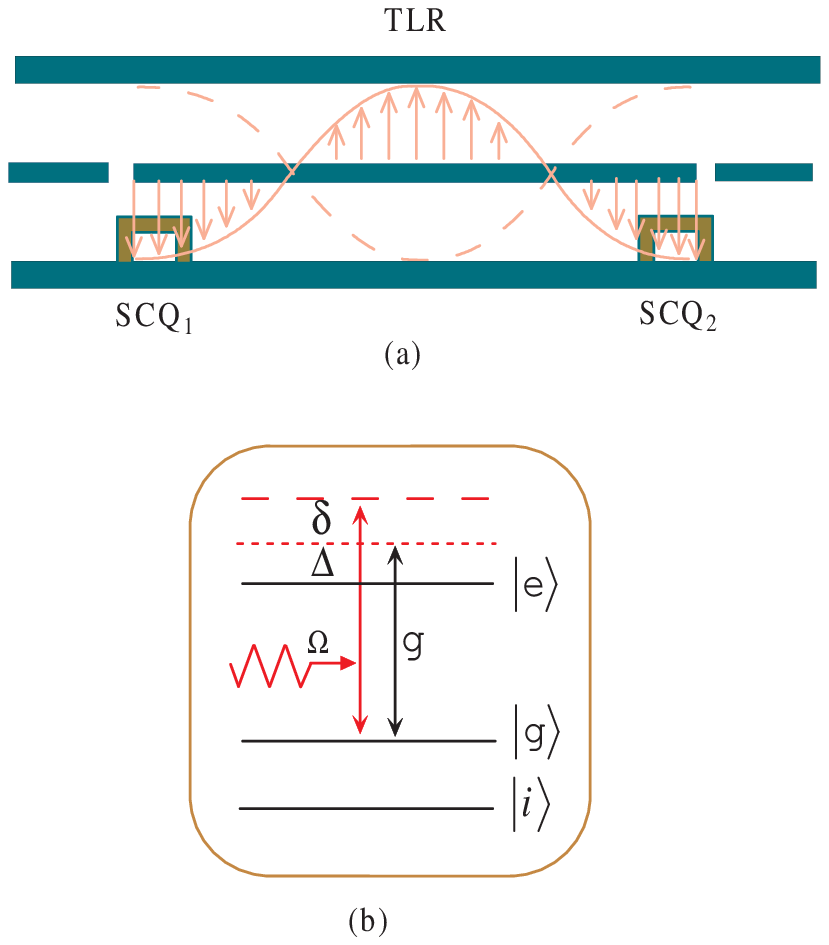}}\caption{\label{f3}}
\end{figure}

\end{document}